\documentclass[aps,prl,superscriptaddress,notitlepage,twocolumn,longbibliography,10pt]{revtex4-2}

\usepackage{amsmath,amssymb,amsthm,bm}
\usepackage{graphicx}
\usepackage[colorinlistoftodos]{todonotes}
\usepackage[inline]{enumitem}
\definecolor{darkblue}{rgb}{0.1,0.2,0.6}
\definecolor{darkred}{rgb}{0.8,0.1,0.2}
\definecolor{crimson}{RGB}{164,16,52}
\definecolor{darkgreen}{rgb}{0.31,0.62,0.24}
\usepackage[colorlinks=true, allcolors=crimson]{hyperref}
\usepackage{diagbox}
\usepackage{epigraph}
\usepackage{float}
\usepackage{multirow}
\usepackage{tikz}
\usepackage{mathtools}
\usepackage{braket}
\usepackage{soul} 
\usepackage[all]{xy}
\usepackage{lipsum}
\usepackage[export]{adjustbox}

\usepackage{bbm}

\newcommand{\Rom}[1]{\uppercase\expandafter{\romannumeral#1}}

\newcommand{\mc}{\mathcal}

\makeatletter
\renewcommand*\env@matrix[1][*\c@MaxMatrixCols c]{%
	\hskip -\arraycolsep
	\let\@ifnextchar\new@ifnextchar
	\array{#1}}
\makeatother

\newtheorem*{claim*}{Claim}

\usepackage[normalem]{ulem}

\definecolor{darkred}{rgb}{0.8,0.1,0.2}

\newcommand*{\ShortSecTitle}[1]{\textit{#1} ---}

\newcommand{\KWI}{U_{1d}}
\newcommand{\tKWI}{\widetilde U_{1d}}
\newcommand{\KWII}{U_{2d}}

\begin{document}
\title{Shallow Unitary Circuits for Kramers-Wannier Dualities}
\author{Yanting Cheng}
\affiliation{Institute of Theoretical Physics and Department of Physics, University of Science and Technology Beijing, Beijing 100083, China}
\author{Shang Liu}
\email{sliu.phys@gmail.com}
\affiliation{Institute of Physics,
Chinese Academy of Sciences, Beijing 100910, China}

\begin{abstract}
The quantum Kramers-Wannier (KW) duality is a fundamental transformation mapping short-range entangled (SRE) states to long-range entangled (LRE) states. While spatially local unitary circuits require linear-in-system-size depth to implement this duality, the ultimate speed limit for purely unitary circuits equipped with nonlocal connectivity remains an open question. Here, we explicitly construct logarithmic depth, spatially nonlocal unitary circuits that realize the exact $\mathbb{Z}_2$ KW dualities in both one and two spatial dimensions. We further generalize the construction to arbitrary $\mathbb{Z}_n$ KW dualities. Unlike algorithms tailored to prepare specific target states, our circuits implement complete duality maps. Within the symmetric (charge-neutral) sector, these dualities exactly transform arbitrary non-fixed-point SRE states into their corresponding LRE duals. Consequently, our results establish an efficient, purely coherent pathway for exploring phase transitions and topological dualities on modern quantum platforms. 
\end{abstract}

\maketitle
\ShortSecTitle{Introduction}
The classification and preparation of quantum phases of matter is a central pursuit in modern quantum many-body physics. A powerful theoretical tool in this endeavor is the Kramers-Wannier (KW) duality \cite{Vijay2016XCube,Kogut1979LGTReview},
which may be regarded as the gauging of a discrete symmetry and establishes a local equivalence between distinct physical systems. Most notably, the KW duality maps short-range entangled (SRE) states, such as the ground state of a paramagnet, to long-range entangled (LRE) states, such as the Greenberger-Horne-Zeilinger (GHZ) state and topologically ordered states. Realizing this duality transformation physically as a quantum circuit is therefore highly desirable, as it provides a direct route to prepare LRE states and simulate the exotic excitations above them by mapping directly from simpler, trivial SRE states.

The circuit complexity of implementing the KW duality is dictated by the allowed quantum resources. Using spatially local unitary gates, the Lieb-Robinson bound dictates that generating an LRE state from an SRE state requires a circuit depth that scales linearly with the system size \cite{Bravyi2006LRBound,Chen2024Sequential}. Recently, it was shown that by expanding the toolkit to include mid-circuit measurements and classical feed-forward, the KW duality can be realized in strictly constant depth, 
$O(1)$ \cite{Nat2021KWfromMeasurement,Bravyi2022PrepareSolvableQD,Nat2022Hierachy,Verstraete2023GaugingCircuit}. At the same time, the historical constraint of strict spatial locality for unitary circuits is being lifted by modern quantum platforms -- such as dynamically reconfigurable Rydberg atom arrays \cite{Antoine2016Science,Antoine2016Nature,Lukin2017Nature,Lukin2023Nature,Antoine2024PRApplied,Lukin2024Nature,Zhou2024NP,Lukin2026Nature} and trapped ions \cite{QCCD2021,TI2023PRX} -- which natively support spatially nonlocal interactions. With current technology on these platforms, executing nonlocal unitary gates is often relatively less resource-intensive than performing fast, non-destructive mid-circuit measurements and real-time classical feed-forward. 
This evolving hardware reality prompts a fundamental question: Can we efficiently realize the KW duality using purely unitary, spatially nonlocal circuits, and what is the ultimate speed limit for such a transformation? Existing no-go results dictate that even with all-to-all connectivity, purely unitary circuits cannot realize the KW duality in $O(1)$ depth, as neither the GHZ state \cite{Aharonov1998MixedStates} nor toric code ground states \cite{Stephen2022NonlocalCircuit,Aharonov2018CircuitDepth} can be prepared with $O(1)$-depth unitary circuits. This leaves the optimal unitary compilation of the KW duality an open problem.

In this Letter, we bridge this gap by explicitly constructing logarithmic-depth unitary circuits that implement exact KW duality maps. Specifically, we first show that the $\mathbb{Z}_2$ KW duality in both one and two spatial dimensions ($1d$ and $2d$) can be realized by a unitary circuit of depth $O(\log_2 N)$ (where $N$ is the total number of qubits) using at-most-two-qubit gates with non-local spatial connectivity. We then extend the result to arbitrary $\mathbb{Z}_n$ KW dualities.

Our results have immediate practical and fundamental implications. While highly optimized $O(\log N)$-depth circuits already exist for preparing specific target LRE states (such as the ideal GHZ or toric code states) \cite{IBM2018GHZExperiment,MooreNilsson1998Parallel,Vidal2008MERA,Vidal2008QuantumDouble,Vidal2009StringNet,White2016FreeFermionMERA,Scholz2018FreeFermionMERA}, our circuits implement the complete duality maps. This means our circuits are not restricted to specific states with nice analytical structures; they can map arbitrary non-fixed-point SRE states (within the charge-neutral sector) to non-fixed-point LRE states, preserving finite correlation lengths and transforming local excitations into anyons. Consequently, our protocol provides an efficient, purely unitary pathway to explore the rich physics of topological matter, phase transitions, and dualities on modern quantum hardware equipped with nonlocal connectivity.

\ShortSecTitle{KW Duality in 1d}
Consider a $1d$ periodic chain of qubits labeled by $j=1,2,3,\cdots,N$. The $\mathbb{Z}_2$ KW duality is a linear transformation under which \footnote{We say that an operator $O$ transforms as $O\mapsto O'$ under a linear map $f$ if $fO=O'f$. } $(Z_jZ_{j+1},X_j)\mapsto (X_{j+1},Z_jZ_{j+1})$. 
Our goal is to construct a shallow unitary circuit realizing this duality. Unitary gates in the circuit should all be single- or two-qubit gates, but are allowed to be spatially nonlocal. 

With periodic boundary condition (PBC), the operator mapping $(Z_jZ_{j+1},X_j)\mapsto (X_{j+1},Z_jZ_{j+1})$ would imply $\prod_j X_j\mapsto 1$, and hence can \emph{not} be realized as a unitary transformation on the whole Hilbert space. For this reason, people often define the KW duality as a noninvertible operator that annihilates all $\mathbb{Z}_2$ odd states, i.e. states with $-1$ eigenvalue under $\prod_j X_j$. Here, we will instead construct a \emph{unitary} KW duality map denoted as $\KWI$, which realizes the desired operator transformation $(Z_jZ_{j+1},X_j)\mapsto (X_{j+1},Z_jZ_{j+1})$ \emph{when acting on $\mathbb{Z}_2$ even states}. More explicitly, under the conjugation of $\KWI$, 
\begin{align}
    X_j&\mapsto Z_jZ_{j+1}\quad (1\leq j\leq N-1), \label{eq:KW1Xj}\\
    X_N&\mapsto Z_N Z_1(\prod_jX_j),\label{eq:KW1XN}\\
    Z_{j-1}Z_{j}&\mapsto X_j\quad (1\leq j\leq N-1), \\
    Z_{N-1}Z_N&\mapsto X_N(\prod_jX_j). 
\end{align}
We observe that near the boundary, the desired KW duality map for PBC has been dressed by the global operator $\prod_jX_j$, which generates a ``dual'' $\mathbb{Z}_2$ symmetry. This dressing can be eliminated if we restrict to $\mathbb{Z}_2$ even initial states: Suppose the initial state $\ket{\psi}$ statisfies $\prod_jX_j\ket{\psi}=\ket{\psi}$. Since $\KWI(\prod_jX_j)\KWI^\dagger=\prod_jX_j$, the final state $\ket{\psi'}:=\KWI\ket{\psi}$ will also be $\mathbb{Z}_2$ even, i.e. $\prod_jX_j$ is equivalent to the identity operator when acting on $\ket{\psi'}$. The above operator mapping rule then reduces to the desired form. In particular, $(X_N,Z_{N-1}Z_N)\mapsto (Z_NZ_1,X_N)$.  

As an application, consider the transverse-field Ising Hamiltonian $H_\text{Ising}(J,h)=-J\sum_jZ_jZ_{j+1}-h\sum_j X_j$ with PBC ($Z_{N+1}\equiv Z_1$) and a $\mathbb{Z}_2$ symmetry generated by $\prod_jX_j$. If $\ket{\psi}$ is a $\mathbb{Z}_2$ even eigenstate of $H_\text{Ising}(J,h)$, then $\KWI\ket{\psi}$ will be a $\mathbb{Z}_2$ even eigenstate of the dual Ising model $H_\text{Ising}(h,J)$ with the same energy. For example, if $\ket{\psi}$ is the product state $\ket{+}^{\otimes N}$, then $\KWI\ket{\psi}$ will be the long-range entangled GHZ state $(\ket{0}^{\otimes N}+\ket{1}^{\otimes N})/\sqrt{2}$. 

For completeness, we also write down the transformation of a single Pauli $Z$ under $\KWI$: 
\begin{align}
    Z_j &\mapsto X_1X_2\cdots X_j Z_N \quad (1\leq j\leq N-1),\label{eq:KW1Zj}\\
    Z_N &\mapsto Z_N. \label{eq:KW1ZN} 
\end{align}
Eqs.\,\ref{eq:KW1Xj}, \ref{eq:KW1XN}, \ref{eq:KW1Zj}, and \ref{eq:KW1ZN} uniquely determine $U_{1d}$ up to an unimportant overall phase. 

We now construct a shallow unitary circuit that implements the KW duality in $1d$. Before presenting the construction, it is useful to fix the notation. We denote by $\KWI(M)$ the KW duality circuit on a chain of $M$ sites, and by $\tKWI(M)$ the \emph{locally rotated} KW duality defined as $\KWI(M)=\left(\prod_{j=1}^{M}\mathcal H_j\right)\tKWI(M)$, where $\mathcal H$ represents the Hadamard gate. Throughout this work, when we say an operator $O$ transforms as $O\mapsto O'$ under a linear map $f$, the precise meaning is $fO=O'f$ as already mentioned in a previous footnote. We also adopt the convention that $\mathrm{CNOT}_{i,j}$ denotes a CNOT gate with control qubit $i$ and target qubit $j$. For instance,
\begin{align}
    &{\rm CNOT}_{1,2}(X_1,Z_1,X_2,Z_2){\rm CNOT}_{1,2}\nonumber\\
    &=(X_1X_2,Z_1,X_2,Z_1Z_2). 
\end{align}
For the sake of simplicity, let us assume the site number $N=2^p$ for some integer $p$. We will comment on the general situation later on. 

\begin{figure}
    \centering
    \includegraphics{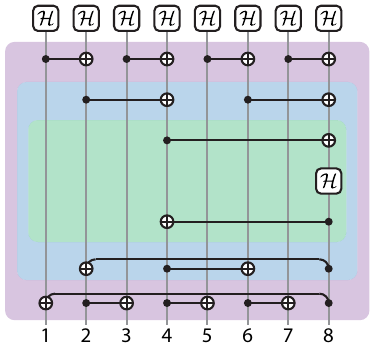}
    \caption{The $1d$ KW duality circuit for $N=8$, with circuit depth $2\log_2N+2=8$. Time goes up. The colored blocks highlight the recursive structure of the circuit.}
    \label{fig:1dKWCircuit}
\end{figure}

We will adopt an inductive construction of the rotated KW duality circuit $\tKWI$, from which one can generate the more standard duality $\KWI$ by an additional layer of Hadamard gates. Suppose that the rotated KW duality has already been implemented on $M$ number of sites, $\tKWI(M)$. We claim that $\tKWI(2M)$ acting on $2M$ sites can be implemented by the following three steps: (1) Apply $V_1(2M):=\prod_{k=1}^{M}\mathrm{CNOT}_{2k,2k+1}$, (2) apply $\tKWI(M)$ to the \emph{even sublattice}, namely the sites with $j=2,4,6,\cdots, 2M$, and finally (3) apply $V_2(2M):=\prod_{k=1}^{M}\mathrm{CNOT}_{2k-1,2k}$. In a compact form, 
\begin{align}
    \tKWI(2M)=V_2(2M)\left[\underset{\color{blue}\text{even}}{\tKWI(M)}\otimes\underset{\color{blue}\text{odd}}{I^{\otimes M}}\right]V_1(2M). 
\end{align}
To initialize the recursion, we choose the minimal building block $\tKWI(1)$ to be a Hadamard gate. 
This is equivalent to $\KWI(1)=I$, consistent with Eqs.\,\ref{eq:KW1XN} and $\ref{eq:KW1ZN}$. 
Starting from this seed, one may verify recursively that the above construction realizes the desired KW transformation for arbitrarily large $N=2^p$. Since each recursion step adds two parallel CNOT layers, the circuit has logarithmic depth $O(\log_2 N)$. For illustration, the circuit $\KWI(8)$ realizing the KW duality on an eight-site chain is shown in Fig.~\ref{fig:1dKWCircuit}.

\begin{figure*}
    \centering
    \includegraphics{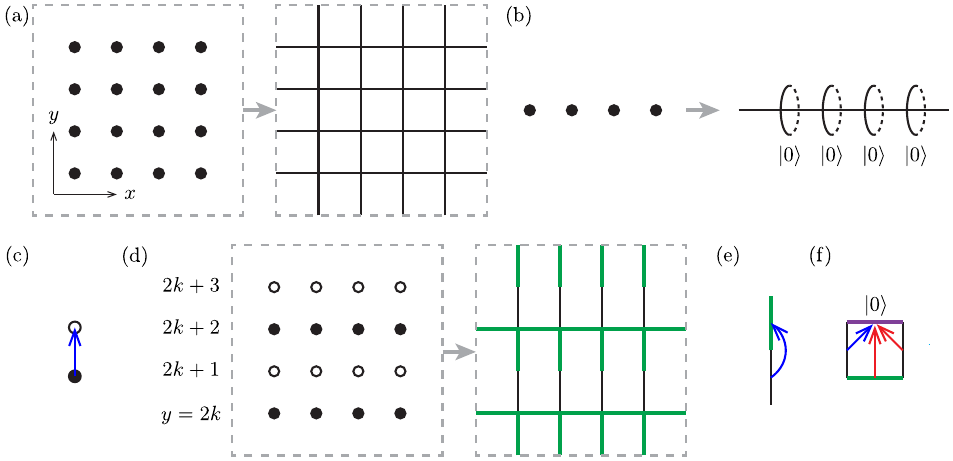}
    \caption{The realization of two-dimensional $\mathbb{Z}_2$ KW duality. (a) The lattice geometry of the duality map. (b) The special case of an $N_1\times 1$ square lattice. (c)-(f) are illustrations of $V_1$, $U$, $V_2$, and $V_3$,  respectively, which constitute one recursive step in our construction of the duality circuit. }
    \label{fig:2dKWCircuit}
\end{figure*}

Let us now take care of more general values of $N$. Suppose $2^p<N<2^{p+1}$ for some positive integer $p$. Let $M=2^p$ and $M'=N-2^p<M$. We may first generate $\tKWI(M)$ by the same recursive procedure. The remaining task is to further extend the site number from $M$ to $M+M'=N$. To this end, we split the $N$ qubits into two sublattices $A$ and $B$ with $|A|=M$ and $|B|=M'$. We require that sites in $B$ are not adjacent, and that the last site $j=N$ belongs to $A$. For example, we can choose $B$ to contain the sites with $j=N-1,N-3,\cdots,N-2M'+1$. One can check that
\begin{align}
    \tKWI(2M)=V_2(N)\left[\underset{\color{blue}A}{\tKWI(M)}\otimes\underset{\color{blue}B}{I^{\otimes M'}}\right]V_1(N), 
\end{align}
where $\tKWI(M)$ acts on the $A$ sublattice, and
\begin{align}
    V_1(N)&:=\prod_{j\in B}\mathrm{CNOT}_{j-1,j},\\
    V_2(N)&:=\prod_{j\in B}\mathrm{CNOT}_{j,j+1}. 
\end{align}

\ShortSecTitle{KW Duality in 2d}
Next, we will show that the $2d$ $\mathbb{Z}_2$ KW duality can also be realized with a unitary circuit of logarithmic depth. 

Consider a $2d$ square lattice with qubits living on the vertices (sites), as shown in the left-hand side of Fig.\,\ref{fig:2dKWCircuit}a. We label the vertices by integer coordinates $(x,y)$ with $x=1,2,\cdots, N_1$, $y=1,2,\cdots, N_2$, and assume the PBC: $x\sim x+N_1$, $y\sim y+N_2$. The $2d$ KW duality is a linear transformation from this Hilbert space to a \emph{distinct} Hilbert space where qubits live on the edges (links) of the same square lattice; see the right-hand side of Fig.\,\ref{fig:2dKWCircuit}a. The operator transformation rule is as follows: (1) For each pair of vertices $(v,v')$ connected by an edge $e$, $Z_vZ_{v'}\mapsto Z_e$. (2) $X_v\mapsto \prod_{e\supset v}X_e$ where the product is over the four edges containing the vertex $v$. Analogous to the $1d$ case, the above operator mapping can \emph{not} be realized by an isometry acting on the whole vertex Hilbert space, because the second rule would imply $\prod_vX_v\mapsto 1$ and lead to a contradiction. We will henceforth \emph{restrict to $\mathbb{Z}_2$ even initial states}, where the $\mathbb{Z}_2$ symmetry is generated by $\prod_vX_v$. The duality circuit $\KWII$ to be constructed below effectively realizes the above operator transformation when acting on such initial states. 

As a consequence of the first rule, $Z_vZ_{v'}\mapsto Z_e$, states in the image of the KW duality map (assuming $\mathbb{Z}_2$ even initial states) should satisfy the ``flux-free'' condition: Given any loop $l$ formed by the edges of the lattice, $\prod_{e\in l}Z_e=1$. As an application, consider the paramagnetic initial state $\ket{\psi}=\prod_v\ket{+}_v$, which satisfies $X_v\ket{\psi}=\ket{\psi}$ for all $v$ and is in particular $\mathbb{Z}_2$ even. The dual state $\ket{\psi'}:=\KWII\ket{\psi}$ will satisfy both $\prod_{e\supset v}X_e\ket{\psi'}=\ket{\psi'}$ and $\prod_{e\subset f}Z_e\ket{\psi'}=\ket{\psi'}$, where the second product is over the four edges of an arbitrary face (plaquette) $f$. These imply that $\ket{\psi'}$ is a toric code ground state. 

The $2d$ KW duality circuit will again be constructed by induction. We assume $N_2=2^q$ for some integer $q$; the general situation can be discussed in a similar way as in $1d$ but will be skipped for the sake of simplicity. 

Suppose we have realized the KW duality on an $N_1\times M$ square lattice. We now proceed to construct a duality circuit for an $N_1\times 2M$ lattice. We start by defining an isometry $U$, acting on $N_1\times 2M$ vertices, which implements the $2d$ KW duality on all qubits with \emph{even} $y$ coordinates. A picture of this isometry is given by Fig.\,\ref{fig:2dKWCircuit}d and the more precise definition is as follows: (1) Black dots (representing qubits with even $y$ coordinates) are mapped to thick, green edges by the KW duality. Each $X$ on a black vertex is mapped to the product of four $X$'s on the four surrounding green edges. Each pair of neighboring $Z$'s on the black sublattice (either horizontally or vertically separated) are mapped to a single $Z$ acting on the green edge in between. (2) Each white dot (representing a qubit with an odd $y$ coordinate) is identified with the thin, black edge right below it; this is purely a pictorial reinterpretation and no actual transformation has been applied to the white dots. The isometry $U$ can be upgraded to the KW duality on the whole $N_1\times 2M$ lattice by combining with three additional isometries: (1) The first isometry/unitary $V_1$, to be applied \emph{before} $U$, is defined as $\prod_{x,k}{\rm CNOT}_{(x,2k),(x,2k+1)}$. A single component of $V_1$ is illustrated in Fig.\,\ref{fig:2dKWCircuit}c, where an arrow $i\rightarrow j$ represents ${\rm CNOT}_{i,j}$. (2) The second isometry/unitary $V_2$, to be applied \emph{after} $U$, consists of ${\rm CNOT}$ gates between vertical edges as shown in Fig.\,\ref{fig:2dKWCircuit}e. (3) The third isometry $V_3$ is to be applied \emph{after} $V_2$. It introduces ancilla qubits initialized in $\ket{0}$, represented by horizontal purple edges at odd $y$ coordinates, and then applies the ${\rm CNOT}$ gates shown in Fig.\,\ref{fig:2dKWCircuit}f (ignoring the coloring of the arrows for now). $V_3$ in particular ensures the flux-free condition $\prod_{e\subset f}Z_e=1$ on every face (plaquette) $f$ of the type shown in the figure. One can check that the combined map $V_3V_2UV_1$ realizes the desired KW duality map. Importantly, the condition of $\mathbb{Z}_2$ even initial states is consistently preserved by this inductive construction: If $\ket{\psi}$ satisfies $\prod_{x,y}X_{(x,y)}\ket{\psi}=\ket{\psi}$, the state $V_1\ket{\psi}$ will satisfy $\prod_{x,k}X_{(x,2k)}(V_1\ket{\psi})=V_1\ket{\psi}$, which is exactly the right initial state condition for $U$. 

It remains to construct the KW duality circuit for the special case of an $N_1\times 1$ square lattice so as to complete our inductive definition. This special case is illustrated in Fig.\,\ref{fig:2dKWCircuit}b. Directly applying the $2d$ KW duality to this special lattice implies the following operator transformation: (1) $X_x\mapsto X_{x-1/2}X_{x+1/2}$ where $x\in\{1,2,\cdots N_1\}$ labels a vertex and the half-integer indices $x\pm 1/2$ label its neighboring horizontal edges. (2) $Z_xZ_{x+1}\mapsto Z_{x+1/2}$.  
These are nothing but the $1d$ rotated KW duality! What should we do for the vertical edges which all become small loops by the PBC? The flux-free condition dictates that they should all be in the $\ket{0}$ state. Therefore, the $2d$ KW duality for the special case of $N_2=1$ can be realized as follows: Use the $1d$ rotated KW duality circuit $\tKWI$ to map the $N_1$ vertices to horizontal edges, and then add $N_1$ ancilla qubits initialized in the $\ket{0}$ state as vertical edges. One may directly check that the aforementioned induction procedure can indeed be applied to this seed to obtain the duality circuit for $N_2=2$. Further implementing the procedure can generate the duality circuit $\KWII$ for arbitrarily large $N_2$. The overall circuit depth is $O(\log_2N)$ with $N=N_1N_2$ being the total number of vertices of the square lattice. 

\ShortSecTitle{Generalization to $\mathbb{Z}_n$ KW Dualities} 
We have thus far focused on $\mathbb{Z}_2$ KW dualities, which preserve the locality of $\mathbb{Z}_2$ symmetric local operators. In this section, we will extend our results to $\mathbb{Z}_n$ KW dualities with an arbitrary integer $n>2$, constructing analogous logarithmic depth duality circuits. 

We begin by introducing some notations of $\mathbb{Z}_n$ Pauli calculus. In the discussion below, the local Hilbert space at every site will be $n$-dimensional, dubbed a $\mathbb{Z}_n$ spin. Choosing an orthonormal basis $\{ \ket{a}|a\in \mathbb{Z}_n \}$ as the computational basis for a $\mathbb{Z}_n$ spin, we define the generalized Pauli $Z$ operator as $Z:=\sum_{a=0}^{n-1}\omega^n\ket{a}\bra{a}$ where $\omega:=\exp(2\pi i/n)$. The generalized Pauli $X$ operator is defined by $X\ket{a}=\ket{a+1}$, whose eigenstates are $\ket{\phi_k}:=(1/\sqrt{n})\sum_a\omega^{-ka}\ket{a}$ with the eigenvalues $\omega^k$, respectively. It is not hard to verify that $Z^n=X^n=I$ and $ZX=\omega XZ$. The analog of Hadamard gate is the Fourier transform operator $\mc F:=\sum_a\ket{a}\bra{\phi_a}$, whose action on Pauli operators is $\mc F(X,Z)\mc F^\dagger=(Z,X^\dagger)$. The analog of $\mathrm{CNOT}$ gate is the controlled-$X$ operator acting on two sites: $\mathrm{CX}_{1,2}:=\sum_a\ket{a}\bra{a}_1\otimes X^a_2$. $\mathrm{CX}_{1,2}$ commutes with $Z_1$ and $X_2$, and its nontrivial action is
\begin{align}
    \mathrm{CX}_{1,2}^a(X_1,Z_2)\mathrm{CX}_{1,2}^{-a}=(X_1X_2^a,Z_1^{-a}Z_2). 
\end{align}

In one spatial dimension, consider a periodic chain of $\mathbb{Z}_n$ spins labeled by $j=1,2,\cdots, N$, and the $\mathbb{Z}_n$ symmetry generated by $\prod_jX_j$. The operator transformation of the $\mathbb{Z}_n$ KW duality is
\begin{align}
    (Z_j^\dagger Z_{j+1},X_j)\mapsto (X_{j+1},Z^\dagger_jZ_{j+1}). 
    \label{eq:1dZnKW}
\end{align}
The shallow circuit realizing this duality follows directly from the $\mathbb Z_2$ construction through the following gate-by-gate substitutions: (1) Replace every $\mathrm{CNOT}_{i,j}$ gate by a $\mathrm{CX}^\dagger_{i,j}=\mathrm{CX}^{-1}_{i,j}$ gate. (2) Replace the single intermediate-layer Hadamard gate by $\mc F$. (3) Replace the final layer of $N$ Hadamard gates by $\mc F^{-1}$ gates. 
The precise operator transformation thus obtained reads as follows: 
\begin{align}
    X_j&\mapsto Z_j^\dagger Z_{j+1}\quad (1\leq j\leq N-1),\\
    X_N&\mapsto Z_N^\dagger Z_1(\prod_jX_j), \\
    Z_j&\mapsto X_1X_2\cdots X_jZ_N\quad (1\leq j\leq N-1),\\
    Z_N&\mapsto Z_N. 
\end{align}
Based on this, one can verify Eq.\,\ref{eq:1dZnKW} assuming \emph{a charge neutral initial state}. 

In two spatial dimensions, the duality circuit is again very similar to the $\mathbb{Z}_2$ case and its structure is depicted in Fig.\,\ref{fig:2dKWCircuit}. All we need to do is to interpret the blue (red) arrows in Fig.\,\ref{fig:2dKWCircuit} as $\mathrm{CX}^\dagger$ ($\mathrm{CX}$) gates. The effective operator transformation when restricted to charge neutral initial states is as follows: 
\begin{align}
    &Z^\dagger_{(x,y)}Z_{(x+1,y)}\mapsto Z_{(x+1/2,y)}, \\
    &Z^\dagger_{(x,y)}Z_{(x,y+1)}\mapsto Z_{(x,y+1/2)}, \\
    &X_{(x,y)}\mapsto X^\dagger_{(x+1/2,y)}X^\dagger_{(x,y+1/2)}X_{(x-1/2,y)}X_{(x,y-1/2)}. 
\end{align}
Note that operators on the left-hand (right-hand) side act on vertices (edges) of the square lattice. 

\ShortSecTitle{Outlook}
Our results should be generalizable to higher spatial dimensions as well. Nonetheless, it worth mentioning that many topologically ordered states in higher dimensions or with nonlocal interactions can already be constructed from parallelized $\mathbb{Z}_n$ gauging in $1d$ and $2d$ \cite{Williamson2023LayerFracton,Tantivasadakarn2026HGP,Liu2026LayeredGauging}, where the gauging is equivalent to KW dualities up to finite-depth local unitary circuits and ancillae. 
Another interesting future direction is generalizing these nonlocal circuits to the case of an arbitrary nonabelian discrete symmetry -- a task for which even adaptive circuits (allowing for measurements and feedforward) currently lack a general sub-linear depth protocol. 

\ShortSecTitle{Acknowledgments}
We are grateful to Xiao-Qi Sun, Pengfei Zhang, and Wenjun Zhang for helpful discussions. S. L. acknowledges support from the Chinese Academy of Sciences (CAS) under Grant No. YSBR-150 and a startup fund from the Institute of Physics, CAS. Y. C. is supported by NSFC Grant No. 12374251.

\bibliography{Bib_Refs.bib}

\end{document}